\documentclass[reprint,amsmath,amssymb,aps,prl,showpacs]{revtex4-1}
\usepackage{amsmath,amssymb,amsfonts}
\usepackage{graphicx,bm}
\usepackage{slashed}
\usepackage{verbatim}
\usepackage[usenames]{color}
\usepackage{hyperref}

\DeclareGraphicsExtensions{.jpg,.JPG,.jpeg,.pdf,.png,.eps}
\graphicspath{{./graphics/}}

\def\be{\begin{equation}}
\def\ee{\end{equation}}
\def\bea{\begin{eqnarray}}
\def\eea{\end{eqnarray}}
\newcommand{\ba}{\begin{aligned}}
\newcommand{\ea}{\end{aligned}}
\newcommand{\baa}{\begin{array}}
\newcommand{\eaa}{\end{array}}
\def\ben{\begin{enumerate}}
\def\een{\end{enumerate}}

\newcommand{\dagh}{^{\dagger}}

\newcommand\fverb{\setbox\pippobox=\hbox\bgroup\verb}
\newcommand\fverbdo{\egroup\medskip\noindent\fbox{\unhbox\pippobox}\ }
\newcommand\fverbit{\egroup\item[\fbox{\unhbox\pippobox}]}

\newcommand{\la}{\lambda}

\newcommand{\bear}{\begin{eqnarray}}
\newcommand{\eear}{\end{eqnarray}}

\newbox\pippobox

\newcommand{\ov}[1]{\overline{#1}}

\providecommand{\slk}{{\slash\!\!\!\!k}}

\providecommand{\slq}{{\slash\!\!\!\!q}}
\providecommand{\e}{\mathrm{e}}

\relax

\def\6{\partial}

\def\a{\alpha}
\def\nn{\nonumber}

\def\le{\left}
\def\ri{\right}

\def\C0{{\bf C_0}}
\def\Y0{{\bf Y_0}}
\def\G0{{\bf G_0}}
\def\e{\epsilon}
\def\m{\mu}
\def\n{\nu}

\def\s{\sigma}

\def\sq
\def\a{\alpha}

\def\la{\langle}
\def\ra{\rangle}

\def\bz{\begin{itemize}}
\def\ez{\end{itemize}}
\def\bn{\begin{enumerate}}
\def\en{\end{enumerate}}

\def\ben{\begin{enumerate}}
\def\een{\end{enumerate}}

\def\a{\alpha}
\def\o{\omega}
\def\6{\partial}

\def\le{\left}
\def\ri{\right}

\def\be{\begin{equation}}
\def\ee{\end{equation}}
\def\bea{\begin{eqnarray}}
\def\eea{\end{eqnarray}}
\def\bz{\begin{itemize}}
\def\ez{\end{itemize}}
\def\la{\langle}
\def\ra{\rangle}
\def\nn{\nonumber}

\newcommand{\si}{\sigma}

\newcommand{\om}{\omega}

\begin{document}

\title{Electromagnetic response of interacting Weyl semimetals}

\author{V.P.J. Jacobs}
\email{vpj.jacobs@gmail.com}
\author{Panagiotis Betzios}
\email{P.Betzios@uu.nl}
\author{Umut G\"ursoy}
\author{H.T.C. Stoof}
\affiliation{Institute for Theoretical Physics and Center for Extreme Matter and Emergent Phenomena, Utrecht University, Leuvenlaan 4, 3584 CE Utrecht, The Netherlands}
\date{\today}

\begin{abstract} We study the electromagnetic properties of Weyl semimetals with strong interactions. Aiming for a large-$N$ expansion, we induce strong interactions by coupling a Weyl fermion with a tunable coupling constant $g_f$ to a quantum critical system with a large number of order $O(N)$ fermionic and bosonic degrees of freedom. The critical fluctuations are described by a conformal field theory containing also fermionic composite operators with scaling dimension $\Delta$. Employing the methods of the holographic correspondence, we then derive the effective theory of the Weyl fermions in the presence of external electric and magnetic fields in the large-$N$ limit. In particular, we determine their frequency and momentum-dependent anomalous magnetic moment. We also determine the conductivity of the Weyl semimetal including the vertex corrections consistent with the Ward identity. Finally, we connect our construction to the case of Coulomb interactions in Weyl semimetals by tuning the parameters $\Delta \rightarrow 5/2$ and $g_f^2 \rightarrow e/\sqrt{\hbar c\epsilon_0}$.
\end{abstract}
\maketitle

\paragraph{Introduction.---}
Dirac \cite{Fang2012a,Young2012,Wang2012,Wang2013,Borisenko2013,Liu2014,Neupane2014,Chen2014} and Weyl \cite{Wan2011,Zyuzin2012a,Halasz2012,Vazifeh2013,Weng2015,Xu2015e,Huang2015d,Burkov2015} semimetals are recently discovered classes of three-dimensional gapless semiconductors in which electrons behave at low energies as relativistic massless fermions. In both cases the valence and conduction bands touch at isolated points $\vec{k}_0$ in the Brillouin zone, the so-called Dirac or Weyl nodes. Around a Weyl node, the linearly dispersing bands also have a definite chirality. The main distinguishing property of Weyl semimetals is the topological stability of their Weyl nodes. Indeed, the corresponding two-component Weyl Hamiltonian $H = \pm \hbar v_F \vec{\s} \cdot (\vec{k}-\vec{k}_0)$, with $v_F$ the Fermi velocity that is typically of the order of $10^6$ m/s \cite{Neupane2014}, $\vec{\s}$ the Pauli spin matrices, and $\pm$ determining the two possible chiralities, is topologically equivalent to a magnetic monopole in momentum space
\cite{Volovik2003a,Nielsen1983,Zyuzin2012a}. These topological properties lead to interesting phenomena such as the existence of topological surface states with a Fermi arc \cite{Wan2011} and nondissipative transport caused by the underlying momentum-space Berry phase \cite{Aji2012, Zyuzin2012, Son2012, Grushin2012, Behrends2015, Vazifeh2013, Goswami2013, Hosur2012}. Weyl semimetals were experimentally realized recently and the Fermi arcs were observed first in TaAs \cite{Xu2015e} but now also in various other compounds \cite{Xu2015c, Xu2015d,Xu2015g}.

All of the above properties can be understood in terms of single-electron physics, without invoking interactions between the electrons. However, this is not expected to be true for all observables. For instance in graphene, which is a two-dimensional Dirac semimetal, the Fermi velocity of the electrons is strongly renormalized by interactions \cite{Elias2012}. Moreover, interactions may also induce interesting magnetic and superconducting phases. It is thus important to study also the effects of the interactions among the electrons on the physical properties of Weyl semimetals and better understand the subtle interplay between interactions and topology. Some of the effects of Coulomb interactions have already been shown to result in logarithmic corrections to the spectral-weight function and the conductivity \cite{Hosur2012,Isobe2013, Burkov2011a, Giuliani2012,Rosenstein2013}. 
\begin{figure}[t]
\centering
\includegraphics[width=70mm]{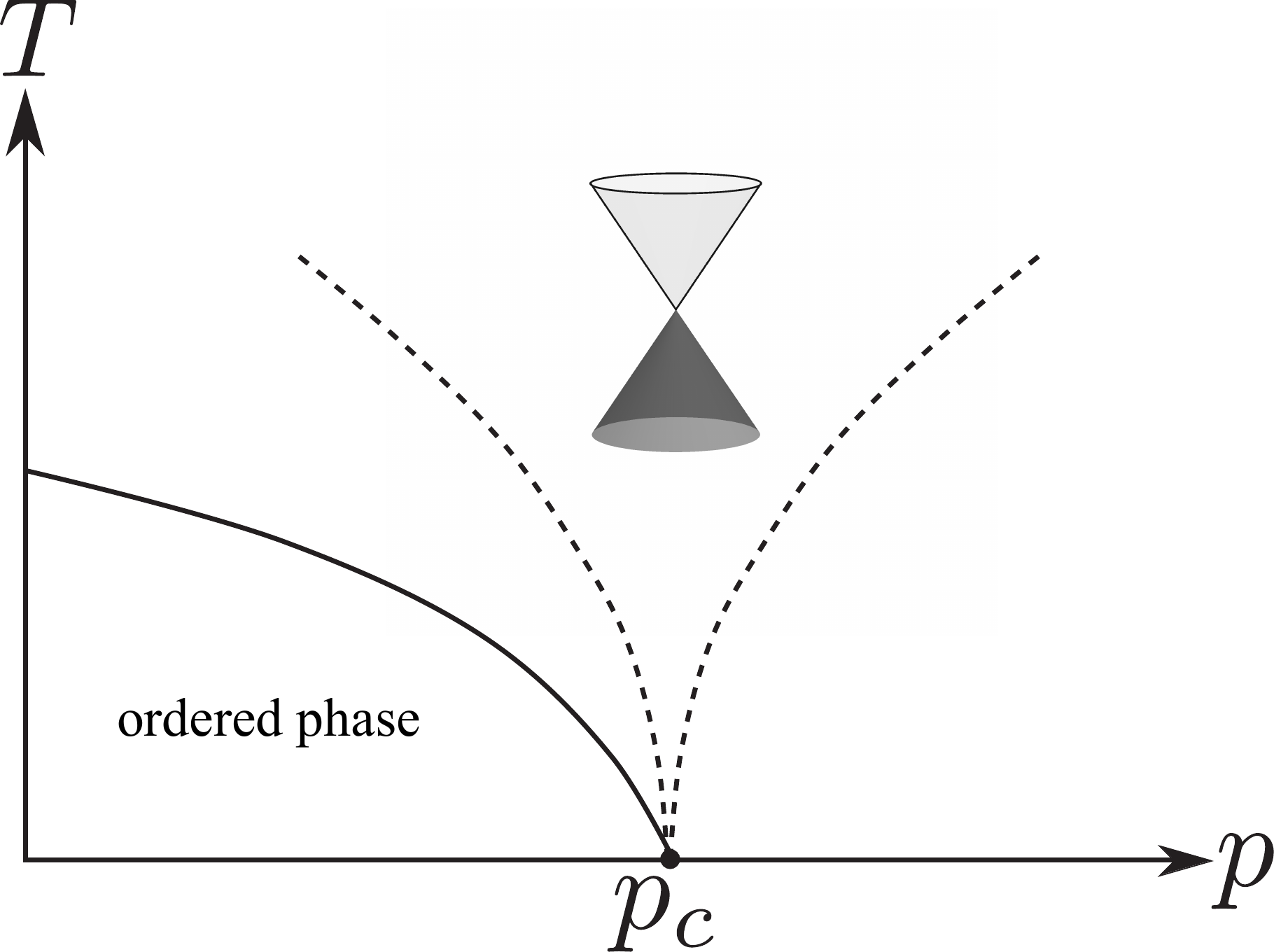}
\caption{Our model describes a charge-neutral Weyl semimetal interacting at zero temperature with a quantum critical system containing fluctuations of both chiralities. The quantum critical region can be reached by tuning a parameter $p$, such as the pressure, to its critical value $p_c$. The strength of the interaction is parameterized by $g_f$ and the universality class of the quantum critical point determines the scaling dimension $\Delta$ of its most relevant fermionic composite operator.}
\label{fig:phasediagram}
\end{figure}

Here, we study the electromagnetic properties of a Weyl semimetal in the presence of a class of long-range interactions at zero temperature and at charge neutrality, where the effects of the interaction are expected to lead to non-Fermi-liquid behavior and the formation of a Weyl liquid. In the spirit of a large-$N$ treatment, we induce these interactions by coupling the Weyl fermions to a strongly interacting quantum critical system with of $O(N)$ fermionic and bosonic degrees of freedom, via the most relevant fermionic composite operator of scaling dimension $\Delta$, as summarized in Fig. \ref{fig:phasediagram}. The universality class of the quantum critical point then determines the parameter $\Delta$ and the associated strongly interacting conformal field theory. The coupling to the conformal field theory is controlled by a tunable coupling constant $g_f$.
We also couple both the chiral electrons and the quantum critical system to an external electromagnetic gauge field.

The main result of this Letter is the effective theory of the Weyl electrons for given external electric and magnetic fields, which requires us to integrate out the quantum critical fluctuations. In general it is a hard task to integrate out a strongly interacting critical system and we manage to do so in the large-$N$ limit by replacing the conformal field theory with its gravitational dual \cite{Maldacena1998, Gubser1998}
\footnote{In the holographic correspondence the conformal field theory is a nonabelian gauge theory having a large number of generators $N_c^2-1 \equiv N \gg1$.}. The resulting effective description incorporates both a selfenergy for the Weyl fermion and corrections to the electromagnetic vertex. The latter allow us to obtain analytic expressions for the charge renormalization and anomalous magnetic moment of the chiral fermion. Finally, we determine the electric response of the Weyl semimetal by calculating its optical conductivity in accordance with the Ward identity for charge conservation.

As a key point, we can connect our results also to the case of Coulomb interactions. Up to logarithmic corrections this corresponds to tuning $\Delta \rightarrow 5/2$ and $g_f^2 \rightarrow e/\sqrt{\hbar c\epsilon_0}$, with $\epsilon_0$ and $c$ respectively the appropriate permittivity and speed of light of the material, since then the critical fluctuations contain a boson with scaling dimension $1$ and hence a propagator that is Coulomb-like \cite{Jacobs2015}. This is explained in more detail below.

\paragraph{Model.---}
Our model is based on the approach of Refs.~\cite{Faulkner2011, Guersoy2013, Guersoy2012}. For definiteness we consider a right-handed Weyl fermion $\chi$ near $\vec{k}_0$ that is minimally coupled to an external gauge field $A_{\mu}$. The Euclidean action is
\be
\ba
S_0[\chi,A]=\hbar \int d k \
\chi^{\dagger}(k)  \le( \frac{i\o}{c} +  \vec{\sigma}\cdot\vec{k}\ri)\chi(k) \\
-e \int  d k\
\chi^{\dagger}(k) \s^\m A_{\m}(q)  \chi(k-q) \, ,
\ea
\ee
where  $\s^\m = (i  \mathbb{I}_2, \vec{\s})$,  $dk= d^4k/(2 \pi)^4$,  $A_{\m}=(A_0 , \vec{A}) = (-i \phi /c, \vec{A})$ with $\phi$ the electric potential, $-e$ is the charge of an electron, and $q$ is the four-momentum of the external gauge field. The chemical potential of $\chi$ is zero at charge neutrality. Note that we have taken $v_F=c$ for simplicity, since this will not affect the long-wavelength physics of the model.
It also contains critical fluctuations that are described by a relativistic and strongly interacting conformal field theory (CFT) with $O(N)$ degrees of freedom and operators of both chiralities. The chiral fermion is coupled to this critical system through a right-handed fermionic operator $\cal O$ in the conformal field theory with scaling dimension $\Delta$
\footnote{In principle $\chi$ can couple to all right-handed CFT operators. For simplicity, we assume $\cal O$ to be the most relevant operator and ignore other possible couplings.}. The action for the conformal field theory and its coupling to the $\chi$-field is
\be
S_1 =g_f \hbar\int dk \left( -i {\cal O}^{\dagger}
\chi + i \chi^{\dagger} {\cal O} \right) + {\cal L}_{\text{CFT}}+\int dk \, J^\mu A_\mu \, ,
\ee
where ${\cal L_{\text{CFT}}}$ is the Lagrangian density and $J^{\m}$ is the charge current of the conformal field theory. The total system $S = S_0 + S_1$ is invariant under $U(1)$ gauge transformations for a conserved total current $J^\m + J^{\m}_{\chi}$, with $J^{\m}_{\chi}=-e\chi^\dagger \sigma^\m \chi$.

Using holographic methods it is possible to integrate out the
strongly interacting CFT \cite{Faulkner2011, Guersoy2013, Guersoy2012}. Anticipating this it is crucial for our purposes to realize that the composite operator ${\cal O}$ can be physically interpreted as the operator $(\sum_{a=1}^N \chi_a \phi_a)/\sqrt{N}$, where $\chi_a$ are chiral fermions and $\phi_a$ are real, possibly composite, scalar bosons. The interaction between the Weyl fermion and the CFT thus resembles the three-point interaction of a large-$N$ theory for strongly interacting electrons with the propagator $\langle \phi_a(x) \phi_{a'}(x') \rangle_{\text{CFT}}$ being the large-$N$ matrix generalization of the (retarded) interaction potential between the electrons. This will indeed correspond to a Coulomb-like interaction if $\phi_a$ has scaling dimension $1$ and thus $\Delta = 3/2 + 1 = 5/2$. Note that in the context of Fig. \ref{fig:phasediagram} the scalar bosons represent the order-parameter fluctuations near the quantum critical point.

\begin{figure}[t]
\centering
\includegraphics[width=80mm]{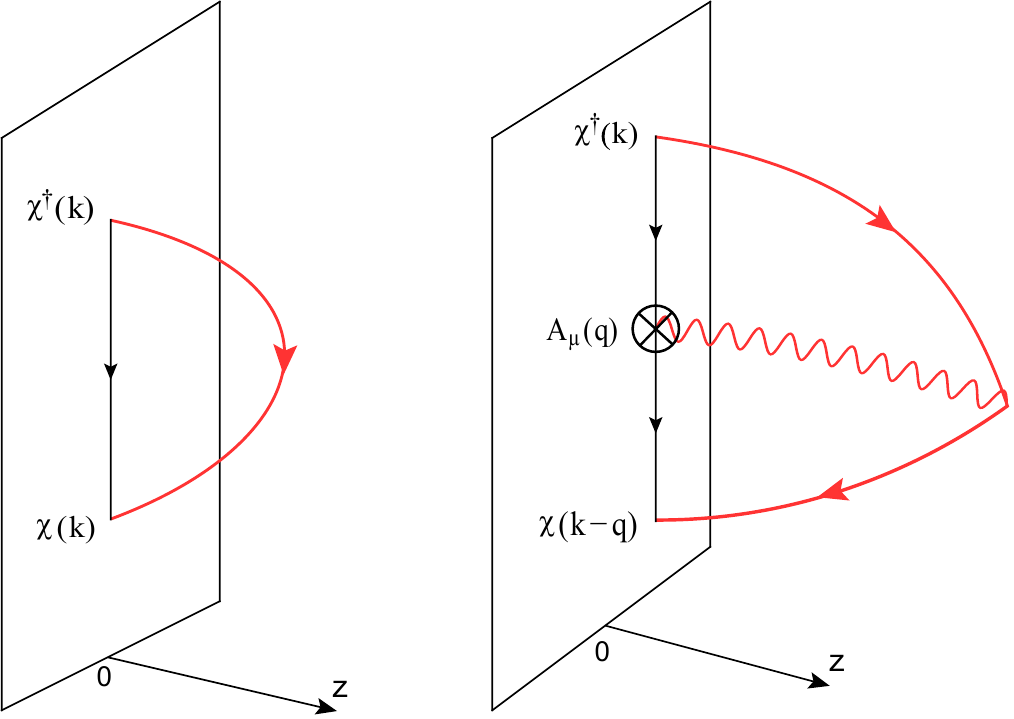}
\caption{The semiholographic Witten diagrams representing the full propagator (left) and the full vertex (right) in the effective action for $\chi$. The effective action is given by a sum over all possible ways of connecting insertions of $\chi$ and $A$ fields through propagation both on the flat boundary $z=0$, and in the curved AdS bulk spacetime $z>0$.}
\label{fig:3ddiagr}
\end{figure}

Taking the large-$N$ limit allows us to integrate out the CFT and results in the semiholographic effective action
\bea
S_{\text{eff}}[\chi, A ] = \phantom{\hspace{-0.7in}} && \\
  && \hbar \int d k\, \chi^{\dagger}(k) \le(\frac{i \o}{c} + \vec{\sigma}\cdot\vec{k} + g_f^2 \la {\cal O}^\dagger{\cal O} \ra_{\text{CFT}}  \ri) \chi(k)  \nn\\
  && + \int d k\, \chi^{\dagger}(k) \left(  -e \s^\m  + g_f^2 \la {\cal O}^{\dagger} J^{\mu} {\cal O} \ra_{\text{CFT}}   \right)A_{\mu} (q)  \chi(k-q).\nn
\eea
This is a linear-response expansion keeping only the terms linear in $A$ as depicted in Fig.~\ref{fig:3ddiagr}. We thus recover that the 2-point function $\langle {\cal O}\dagh{\cal O}\rangle_{\text{CFT}}$ acts as a selfenergy for $\chi$ \cite{Guersoy2012}.
From the above physical interpretation of the operator $\cal O$ this selfenergy can be seen as a selfconsistent Fock diagram, which allows us to make the identification $g_f^2 \rightarrow e/\sqrt{\hbar c\epsilon_0}$ for the Coulomb limit.
A novel fact of this work is that the 3-point correlator $\la {\cal O}^{\dagger}J^{\mu}{\cal O} \ra_{\text{CFT}}$  contributes by dressing the bare electromagnetic vertex. These two CFT correlators are then determined by computing the holographic Witten diagrams dual to them.


\paragraph{Holography.---}At zero temperature and zero chemical potential, the
minimal holographic dual of relativistic, strongly interacting critical fluctuations of both chiralities is a (4+1)-dimensional curved geometry that is known as anti-de Sitter (AdS) space. It has an extra radial coordinate $z$, an AdS radius $\ell$, and an Euclidean metric $g_{MN}$. The correlators of the boundary operators $\cal O$ and $J^{\mu}$ are, respectively, determined from a fermionic field
$\Psi$ of mass $M\hbar/c\ell$ and charge $q_A$, and a gauge field $A_N$ propagating on this curved spacetime background \cite{Mueck1998,
Contino2004}. The associated Euclidean bulk action is
\bea \label{eq:bulkaction} \frac{16\pi G_{\text{N}}}{\hbar} S[\Psi, A]=
\int {\rm d}^{5}x \sqrt{g} \left(
   \ov \Psi \slashed{\mathcal{D}} \Psi
  -\frac{M}{\ell} \,\ov \Psi \Psi \right) \nonumber\\
   +\frac{1}{4} \int {\rm d}^{5}x \sqrt{g} F_{MN} F^{MN}.
\eea
Here, Newton's gravitational constant is denoted by $G_\text{N}$ and $\slashed{\mathcal D}=\slashed{\nabla} - i q_A\, \slashed{A} \sqrt{\ell/ \e_0 \hbar c}$ contains the covariant derivative on AdS and the minimal coupling to the bulk gauge field \footnote{Details of our notation are explained in Ref. \cite{Guersoy2013}.}. The parameter $q_A$ is {\it a priori} arbitrary and will be fixed by the Ward identity below. Finally, the mass $M$ is related to the scaling dimension of the CFT operator ${\cal O}$ as $\Delta = 2+ M$.
Solving the Dirac equation in the bulk results in the 2-point function \cite{Iqbal2009}
\be\label{quad}
\la {\cal O}^\dagger{\cal O} \ra_{\text{CFT}}= \frac{1}{2^{2\Delta-4}} \frac{\Gamma (\frac{5}{2}-\Delta)}{\Gamma (\Delta-\frac{3}{2})} \: \slk \: k^{2\Delta - 5} \equiv \frac{\Sigma(k)}{g_f^2} ,
\ee
with $k^{\mu} = (\o /c , \vec{k})$, $k= \sqrt{k^{\mu} k_{\mu}}$, and $\slk = {k}_\m \s^\m$. Here,
$3/2<\Delta<5/2$ for consistency.

\paragraph{Vertex correction.---} Now we turn to the 3-point correlator that gives the extra contribution to the full electromagnetic vertex $\Lambda^{\mu} \equiv -  e \sigma^\m + g_f^2 \la {\cal O}^{\dagger}J^{\mu}{\cal O} \ra_{\text{CFT}}$. It is given by the 3-point Witten diagram corresponding to the second diagram in Fig.~\ref{fig:3ddiagr}. To evaluate it, we use the bulk-to-boundary propagators for the bulk fermion and gauge field which can be found in Ref.~\cite{Mueck1998}. These depend on the momenta of the particles in the transverse boundary directions, but also on the radial position $z$ of the AdS spacetime. Fig.~\ref{fig:3ddiagr} shows that in the second diagram we also integrate over the radial position of the bulk interaction point. Ultimately, the Witten diagrams can be written in terms of two radial integrals as
\bea
&\la {\cal O}^{\dagger}(k-q)J^{\mu}(q){\cal O}(k) \ra_{\text{CFT}} = \mathcal{N}_\Delta  \Big[ - \sigma^{\mu} I_1(k,k-q,q)   \nn \\ &
 +{\displaystyle  k_{\rho} (k-q)_{\lambda} }~ \sigma^{\rho}\bar{\sigma}^{\m}\sigma^{\lambda} I_2(k,k-q,q)\Big] ,
\eea
\makebox[\linewidth][s]{with $I_{i}=\int dz z^2 k^{\nu_{i}}(k-q)^{\nu_{i}} q K_{\nu_{i}}(kz) K_{\nu_{i}}((k-q)z)K_{1}(qz)$} for $i=1,2$, $\mathcal{N}_\Delta= q_A/2^{2\Delta-5}\Gamma^2(\Delta-{3}/{2})$, and $\bar{\s}^\m = (-i \mathbb{I}_2, \vec{\s})$.
In the case of pure AdS the bulk-to-boundary propagators are proportional to the modified Bessel functions $K_{\nu}(kz)$ with $\nu=1$ for the gauge field. For the fermions we have $\nu_{1} = \Delta-3/2$ and $\nu_{2} = \Delta-5/2$.

The expression above contains all the nontrivial dependence of the vertex on the external momenta \footnote{Generic triple-Bessel integrals can be computed in terms of the Appell hypergeometric $F_4$ function \cite{Prudnikov1998}. In the long-wavelength approximation for the external $A(q)$, we can perform the integral analytically after expanding $K_{1}(qz)$ for $qz \rightarrow 0$. This expansion has been verified by comparing with the exact numerical result.}.
As a check on our computation, we multiply the longitudinal part of the vertex with $q_{\m}$ to relate the 3-point and 2-point correlators by
$q_\m
\la {\cal O}^{\dagger}(k-q)J^{\mu}(q){\cal O}(k) \ra_{\text{CFT}}=  q_A \left[ \la {\cal O}^\dagger{\cal O} \ra_{\text{CFT}}(k-q) -\la {\cal O}^\dagger{\cal O} \ra_{\text{CFT}}(k)  \right]$.
This is the CFT Ward identity \cite{Mueck1998}. Provided this CFT Ward identity holds, the Ward identity for the full vertex $\Lambda^{\mu}$ of the Weyl fermion is also satisfied if $q_A=e$, i.e.,
\bea
&q_\mu \Lambda^\mu (k,k-q)= \hspace{-1pt}-e \  \slq +  q_A g_f^2 \left[ \la {\cal O}^\dagger{\cal O} \ra(k\hspace{-1pt}-\hspace{-1pt}q) -\la {\cal O}^\dagger{\cal O} \ra(k)  \right] \nn\\&
 =e \left[ G^{-1}(k-q) -G^{-1}(k) \right] ,
\eea
with $G(k)=1/(\slashed{k}+\Sigma(k))$.

In addition, the vertex contains a wealth of physical information because of its intricate tensor structure. A prediction of our model coming from the term in the vertex proportional to $\s^{\mu}$ is a frequency and momentum-dependent charge renormalization due to interactions. Even more important is the contribution of the transverse part of the vertex to the effective action, which reads in the long-wavelength limit
\be\label{eq:Sefffinal}
\int dk \chi\dagh(k)\mu_m(k)\left[\sigma^{\rho} k^{\lambda}\left(F_{\rho \lambda}-{}^*F_{\rho \lambda}\right) \right]\chi(k-q). \nonumber
\ee
Here, the dual Faraday tensor is ${}^*F^{\mu \nu}=\epsilon^{\mu \nu \rho \lambda}F_{\rho \lambda}$/2, and $\mu_m(k) = \pi g_f^2\mathcal{N}_\Delta \mbox{$({5}/{2}-\Delta) $} k^{2\Delta-7}/2\cos(\pi(\Delta-2))$ determines the frequency and momentum-dependent magnetic moment that, contrary to the selfenergy, changes its sign for a left-handed Weyl fermion. This interpretation becomes clear if we expand the Faraday tensor into its electric and magnetic components. We then indeed find a Zeeman-like coupling $-i\om \mu_m(k) \vec{\si}\cdot\vec{B}$. The interactions induce also a Stark-like effect given by $-\om \mu_m(k) \vec{\si}\cdot \vec{E}$, a Rashba-like spin-orbit term $- i \mu_m(k) \vec{B}\cdot(\vec{\si}\times \vec{k})$, and an Aharonov-Casher effect from $\mu_m(k) \vec{k}\cdot(\vec{\si}\times\vec{E})$. Note that all these terms are time-reversal symmetric as $\vec{E}$ and $\vec{B}$ denote the physical electric and magnetic fields.

\begin{figure}[t]
\vskip 10pt
\centering
\includegraphics[width=80mm]{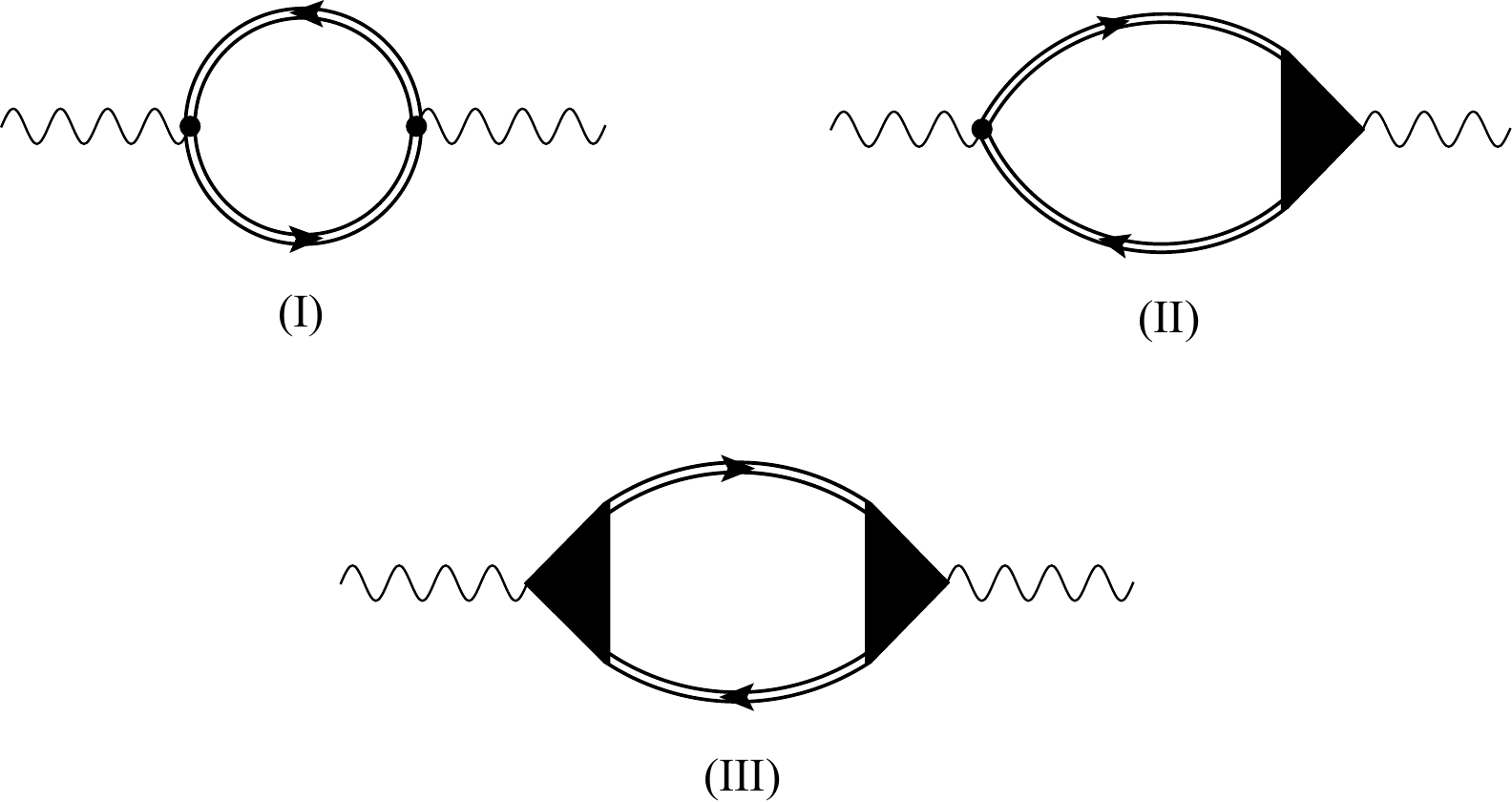}
\caption{Contributions to the current-current correlation function. Double lines denote $G(k)$. The bare vertex diagram $(I)$ is the fermionic contribution $\langle J^{\mu}_{\chi} J^{\nu}_{\chi}  \ra$. The diagrams with one dressed vertex $(II)$ provide the interference term $2 \langle J^{(\mu}_{\chi} J^{\nu)}  \ra$. The diagram with two dressed vertices $(III)$ corrects the CFT current-current correlation function by $\langle J^{\mu} J^{\nu}  \ra-\langle J^{\mu} J^{\nu}  \ra_{\text{CFT}}$.}
\label{fig:Poldiag}
\end{figure}

\begin{figure}[t]
\vskip 10pt
\centering
\includegraphics[width=80mm]{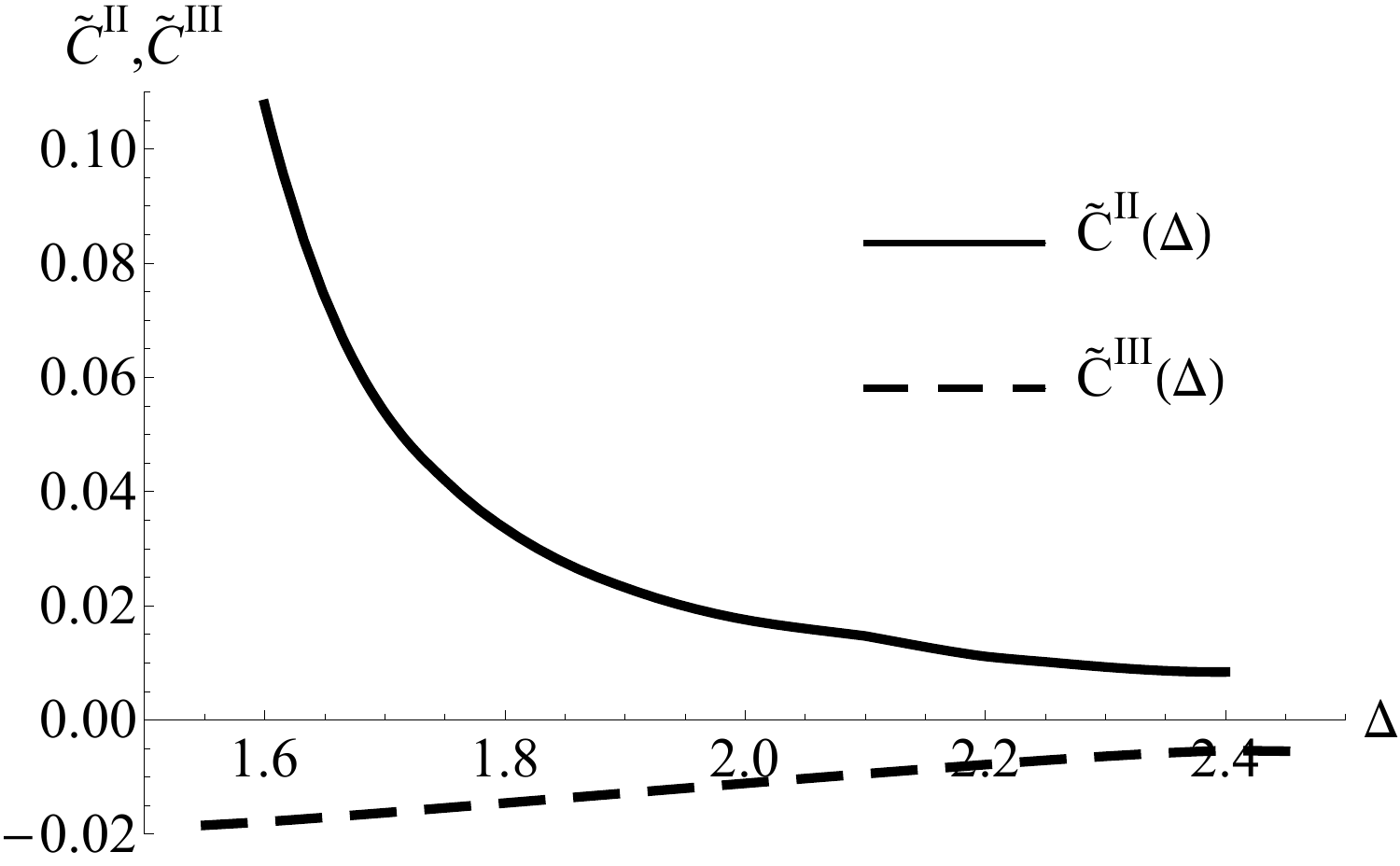}
\caption{The conductivity prefactors as a function of the scaling dimension $\Delta$. We have plotted the dimensionless functions $\tilde{C}^{II}(\Delta)=C^{II}(\Delta,g_f) g_f^2 c^{6-2\Delta}\hbar/e^2$ and $\tilde{C}^{III}(\Delta)=C^{III}(\Delta,g_f) \hbar c /e^2$.}
\label{fig:prefacM}
\end{figure}

\textit{Conductivity.---}The conductivity requires evaluation of the one-loop diagrams depicted in Fig.~\ref{fig:Poldiag}.
This corresponds to computing the current-current correlation function
$\Pi^{\mu \nu}(q)= \int d k \text{Tr}\left[\Lambda^\m G(k) \Lambda^\n G(k-q)\right]$. Applying the Kubo formula $\sigma^{i j}(\omega) =- c\Pi^{i j}(\vec{q}=0, \omega)/\hbar\omega$ the complex optical conductivity can be obtained by a Wick rotation. We are however interested in the real dissipative part of the conductivity and extract this from the imaginary-time result by considering
the nonanalytic part that forms a cusp at $\omega=0$. We focus on the diagonal terms of the conductivity tensor, since the off-diagonal terms are not affected by interactions here, and by virtue of symmetry just on $\sigma^{x x}$. Both the quantum critical system and the Weyl fermion are charged so the total current is $\vec{J}+ \vec{J}_{\chi}$. Thus, we can split the computation of $\Pi^{xx}$ into three parts as in Fig.~\ref{fig:Poldiag}. Using scaling arguments we find the leading behavior as $\om\rightarrow 0$ of these three contributions to the conductivity. From the bare vertex diagram we get $\sigma^{(I)}(\omega)= C^{I}(\Delta, g_f) \omega^{11-4\Delta}$ which can be thought of as the purely fermionic contribution $\langle J^i_{\chi} J^j_{\chi} \ra / \omega$
to the conductivity. The prefactor $C^I$ was already computed previously by a real-time approach in Ref.~\cite{Jacobs2014}. From the diagram with two vertex corrections we find $\sigma^{(III)}(\omega)= C^{III}(\Delta, g_f) \omega$, which provides a correction to the prefactor of the conformal field-theory result $\langle J^i J^j \rangle_{\text{CFT}} / \omega$ that is also linear in $\omega$. This correction is negative because the Weyl fermion leads to additional relaxation of the CFT current. Finally, the two diagrams with one dressed and one bare vertex can be thought of as interference terms between the chiral fermion and the conformal field theory coming from $2 \langle J^i_{\chi} J^j \ra  / \omega$, and evaluate to $\sigma^{(II)}(\omega)= 2C^{II}(\Delta, g_f) \omega^{6-2\Delta}$. We have also computed the full conductivity numerically to check the scaling behavior and determine the coefficients $C(\Delta ,g_f)$. This result is presented in Fig.~\ref{fig:prefacM}. Each of these three diagrams has a different frequency dependence as $\omega \rightarrow 0$, which can be distinguished by experiment.

\paragraph{Discussion.---} At small, nonzero temperature and zero frequency, temperature provides the only energy scale and we obtain for the dc conductivity,
\bea
\sigma_{\text{dc}} = \left[O(N)+O(1)\right]T +O(1) T^{6-2\Delta} + O(1) T^{11-4\Delta} \nn.
\eea
In this expression we show also the order of the number of degrees of freedom. For instance, the $O(N)$ contribution corresponds to $\langle J^{i} J^{j} \ra_{\text{CFT}}$. We calculated here the Weyl-semimetal conductivity up to $O(1)$, whereas $O(1/N)$ corrections come from bulk loop diagrams. For instance, there is a $O(1)$ contribution linear in temperature that originates from the bulk one-loop corrections to the CFT two-point function. In principle, the prefactor of this linear term also receives an $O(1)$ contribution from the quadratic term in the expansion of the selfenergy in $A$ that corresponds to a four-point Witten diagram. These contributions are beyond the scope of the present paper. The full frequency and temperature-dependence of the conductivity can also be determined and we plan to elaborate on both these issues in future work \cite{Betzios2015}.

Very importantly, the $\Delta\rightarrow 5/2$ limit corresponds to Coulomb interactions. In this limit we indeed recover not only the correct scaling behavior, but exactly at $\Delta=5/2$ also logarithmic corrections to both the 2-point correlation function and the conductivity, in agreement with the results of Refs.~\cite{Abrikosov1971,Burkov2011a,Hosur2012,Witczak-Krempa2014}.
We note that the vertex can also be derived using newly developed conformal field-theory techniques \cite{Bzowski2014}. The main advantage of AdS/CFT is that it is easy to explore nonzero temperature and chemical potential by using the appropriate black-hole backgrounds and working out the propagators numerically. It will be very interesting to study our model in this more general setup, also in view of possible application to the physics of quark-gluon plasma's. Moreover, by incorporating the backreaction of an axial gauge field on the geometry we can also study the effects of the separation between Weyl cones on the renormalization of the interactions, that has been neglected thusfar \cite{Hosur2012,Isobe2013, Burkov2011a, Giuliani2012,Rosenstein2013} as this separation does not affect the density fluctuations.
Finally, our work opens the possibility to study several other electromagnetic responses due to the vertex corrections, such as the macroscopic magnetization and polarization of interacting Weyl semimetals with an axial chemical potential, i.e., with a population imbalance in the nodes of opposite chirality, and in the presence of external magnetic and electric fields, which are hard to determine via the usual field-theoretic techniques.

We thank Stefan Vandoren and Jan Zaanen for useful discussions. This work is supported by the Stichting voor Fundamenteel Onderzoek der Materie (FOM) and is part of the D-ITP consortium, a program of the Netherlands Organisation for Scientific Research (NWO) that is funded by the Dutch Ministry of Education, Culture and Science (OCW).


\bibliography{magmoment}

\end{document}